\documentclass{vldb}
\usepackage{graphicx}
\usepackage{listings}
\usepackage{float}
\usepackage{verbatim}
\usepackage{url}
\usepackage{balance}  % for  \balance command ON LAST PAGE  (only there!)

\newcommand{\minisec}[1]{\noindent\textbf{#1}}

\begin{document}
\newbox\abox
\begin{lrbox}{\abox}
\begin{lstlisting}
fun gen_order(w_id, d_id, c_id, order) {
  wh = get("warehouse", w_id);
  dist = get("district", w_id, d_id);
  cust = get("customer", w_id, d_id, c_id);
  add("new_order", w_id, d_id, dist.order_id);
  update("district", "order_id", w_id, d_id, dist.order_id+1);
  add("oorder", w_id, d_id, c_id, dist.order_id, order);
  return <wh,dist,cust>;
}

fun get_amount(ord_item) {
  item = get("item", ord_item.id);
  return item.amount * ord_item.amount;
}

fun update_stock(ord_item, amount) {
  stock = get("stock", ord_item.id, ord_item.supplier_w_id);
  new_stock = f(stock);
  update("stock", item.id, ord_item.supplier_w_id, new_stock, amount, ...);
  return stock;
}
\end{lstlisting}
\end{lrbox}

\newbox\bbox
\begin{lrbox}{\bbox}
\begin{lstlisting}
txn new_order(w_id, d_id, c_id, order) {
   //Get the order id from the district and insert the 
   //next order id to be used in it
  <wh,dist,cust> = gen_order_id(w_id, d_id, c_id, order);

  total = 0;
  for(ord_item in order.items) {
    //Get the amount for the number of items ordered 
    amount = get_amount(ord_item); 
    total += amount;

    //Get the district information of the stock from the 
    //supplier warehouse
    stock_info = get_dist_info_stock(ord_item); 
    
    //Update the stock of the supplier warehouse
    update_stock(ord_item, amount);

    add("order_line", dist.order_id, w_id, d_id, stock_info, amount, ...);
  }
  total_pay = (1 + wh.tax + dist.tax) * total * (1 - cust.discount);
  return total_pay;
}
\end{lstlisting}
\end{lrbox}

\newbox\xbox
\begin{lrbox}{\xbox}
\begin{lstlisting}
txn new_order_update_stock(order) {
  result = <>;
  for(ord_item in order.items) {
    amount = get_amount(ord_item);
    stock_info = get_dist_info_stock(ord_item);
    update_stock(ord_item, amount);
    append(result,<stock_info,amount>);
  }
  return result;
}
\end{lstlisting}
\end{lrbox}

\newbox\ybox
\begin{lrbox}{\ybox}
\begin{lstlisting}
PARTITIONING FUNCTION map(w_id) {return w_id;};
new_order PARTITION MAPPER map;
new_order_update_stock PARTITION MAPPER map;
\end{lstlisting}
\end{lrbox}

\newbox\cbox
\begin{lrbox}{\cbox}
\begin{lstlisting}
txn new_order (w_id, d_id, c_id, order) {
  <wh,dist,cust> = gen_order_id(w_id, d_id, c_id, order);

  //Execute the remote transaction for all the order items 
  //requested of the supplier warehouse in parallel
  PARALLEL EXEC new_order_update_stock 
  (subset(order, s_id)) ON PARTITION (s_id) 
  for s_id in order.supplier_w_id;

  total = 0;
  for(ord_item in order.items) {
    //Compute the pay amount
    amount = get_amount(ord_item);
    total += amount;
    //Get the district information of the stock using the
    //replicated stock table
    stock_info = get_dist_info_stock(ord_item);
    add("order_line", dist.order_id, w_id, d_id, stock_info, amount, ...);
  }
  total_pay = (1+wh.tax+dist.tax)*total*(1-cust.discount);
  return total_pay;
}
\end{lstlisting}
\end{lrbox}

\newbox\dbox
\begin{lrbox}{\dbox}
\begin{lstlisting}
txn new_order (w_id, d_id, c_id, order) {
  <wh,dist,cust> = gen_order_id(w_id, d_id, c_id, order);
  results[order.num_s_w_id]; i=0;

  for(s_id in order.supplier_w_id) {
    //Execute the remote transaction for all the order
    // items requested of the remote supplier warehouse
    results[i++] = EXEC new_order_update_stock 
    (subset(order, s_id)) ON PARTITION (s_id);
  }

  //Use results to add order line and compute total pay
  total = 0;
  for(result in results) {
    for(item_result in result) {
       total += item_result.amount;
       add("order_line", dist.order_id, w_id, d_id, item_result, ...);
    }
  }
  total_pay = (1+wh.tax+dist.tax)*total*(1-cust.discount);
  return total_pay;
}
\end{lstlisting}
\end{lrbox}

% ****************** TITLE ****************************************

\title{Transactional Partitioning: A
New Abstraction for Main-Memory Databases}

\numberofauthors{1} %  in this sample file, there are a *total*

\author{
\alignauthor
Vivek Shah\\
(supervised by Marcos Vaz Salles)\\
       \affaddr{University of Copenhagen}\\
       \affaddr{Copenhagen, Denmark}\\
       \email{bonii@di.ku.dk}
}

\maketitle

\begin{abstract}
The growth in variety and volume of OLTP (Online Transaction
Processing) applications poses a challenge to OLTP systems to
meet performance and cost demands in the existing hardware
landscape. These applications are highly interactive (latency
sensitive) and require update consistency. They target commodity
hardware for deployment and demand scalability in throughput with
increasing clients and data. Currently, OLTP systems used by
these applications provide trade-offs in performance and ease of
development over a variety of applications. In order to bridge
the gap between performance and ease of development, we propose
an intuitive, high-level programming model which allows OLTP
applications to be modeled as a cluster of application logic
units. By extending transactions guaranteeing full ACID semantics
to provide the proposed model, we maintain ease of application
development. The model allows the application developer to reason
about program performance, and to influence it without the
involvement of OLTP system designers (database designers) and/or
DBAs. As a result, the database designer is free to focus on
efficient running of programs to ensure optimal cluster resource
utilization.
\end{abstract}

\section{Introduction}
OLTP applications have grown in volume, variety and performance
demands \cite{DBLP:journals/cacm/Stonebraker12}. In addition to the classic banking, reservation, and
order entry systems, novel applications include massively
multiplayer online worlds (MMOW) or virtual worlds, massively
multiplayer online role playing games (MMORPG), financial
applications (e.g., online trading), telecommunication
applications and information visualizations. These applications
demand scalability in throughput with growing data and
clients. They exhibit high interactivity (latency sensitive)
while maintaining consistency on updates. Furthermore, they
typically target commodity hardware for deployment. We will refer
to them as HICCUP (Highly-Interactive Commodity-Hardware
Consistent-on-Update) applications.

Stonebraker et al.  have shown that OLTP systems can be deployed
in main-memory using shared-nothing cluster of machines at modest
cost \cite{DBLP:conf/vldb/StonebrakerMAHHH07}. This leads to
order-of-magnitude gains in performance over classical relational
database management systems (RDBMS). With the reduction in prices
of commodity hardware and growth of cloud computing
\cite{DBLP:journals/cacm/ArmbrustFGJKKLPRSZ10}, a cluster-based
infrastructure consisting of shared-memory commodity hardware
nodes increasingly provides flexible main-memory deployment
options to meet HICCUP application requirements.

In order to build HICCUP applications for cluster-based
architectures, there are currently two dominant programming
models exposed to the application developer. The first approach
abstracts the distributed architecture into a uniform
shared-memory space, which is then presented using a unified data
model, e.g., the relational model under strong consistency
semantics (ACID transactions). The second approach abstracts the
distributed architecture into a low-level distributed
storage-oriented model, e.g., key value stores with no/loose
consistency semantics.

Although the first approach provides ease of development,
performance varies with the variety of applications. This is
because of the sensitivity of the approach to data and code
partitioning to reduce the impact of distributed transactions
\cite{DBLP:conf/cidr/Helland07}. In the second approach, performance is controlled
by the application developer, but development is extremely hard
due to no/loose consistency semantics and the low-level, 
storage-oriented programming model. There are in-between
approaches that partition the unified data model and provide
consistency semantics within partitions \cite{DBLP:conf/cidr/BakerBCFKLLLLY11,
DBLP:journals/pvldb/ShuteVSHWROLMECRSA13}. However, in these systems, the
clients must specify these partitions and handle consistency
across partitions. This increases application complexity,
especially for HICCUP applications which cannot always be
architected to eliminate inter-partition accesses.

We want to address this gap by exploring a middle ground between
the two extreme approaches. Strong consistency semantics (ACID
transactions) provide ease of application development. Sacrificing
consistency semantics is not enough to guarantee OLTP application
performance \cite{DBLP:journals/pvldb/FloratouTDPZ12}. We want to
build a transactional system maintaining global consistency
semantics in the cluster to enable ease of development while
guaranteeing optimal cluster performance. In order to do so, we
propose to extend the transactional abstraction with a logical
distribution abstraction. This enables the application developers
to reason about the performance of their application in terms of
distributed transactional logic units and thus write
efficient programs.

While the logical distribution abstraction hides the low-level
distributed architecture, it allows application developers to
understand program performance behavior in a cluster setup and to
influence it. Database designers can then focus on running the
programs efficiently for optimal cluster resource
utilization. This approach ensures that the application
developers can participate in improving the performance of the
application. The approach also reduces the criticality of data
and code partitioning. As a result, it removes database designers
from the critical path of performance improvement over a variety
of applications by separating concerns of writing efficient
programs (application developers) and running them efficiently
(database designers). Global consistency guarantees in the
cluster free the developer from worrying about consistency
issues, thus reducing program complexity and development costs.

The remainder of the paper is organized as follows. In Section
\ref{sec:background}, we provide a background of the project by
laying down the design requirements for OLTP systems and
reviewing the state of the art with respect to these
requirements. In Section \ref{sec:approach}, we describe our
approach in greater detail with an example OLTP application (the
TPC-C
\footnote{\label{footnote:tpcc}\url{http://www.tpc.org/tpcc/spec/tpcc_current.pdf}}
benchmark). Finally, in Section \ref{sec:roadmap}, we outline the
challenges and roadmap over the course of the project with some
initial thoughts on how to tackle them.

\section{Background}
\label{sec:background}
In this section, we provide a background of the Ph.D. project by
first isolating the OLTP design requirements in the current
hardware and software landscape and then reviewing the state of
the art with respect to these design requirements. 

\subsection{OLTP Design Requirements}
\begin{enumerate}
\item In order to meet the requirements of a growing variety of
HICCUP applications, OLTP systems must expose a programming model
which allows application developers to reason about program
performance in a cluster-based setup and to improve the performance of
their architected applications (programs). 
\item Lack of global consistency semantics and a low-level
 distributed storage-oriented programming model make application
development extremely expensive, error-prone and limit the
adoptability of such a system
\cite{DBLP:journals/cacm/Terry13}. OLTP systems must provide
a high-level, intuitive programming model with strong global
consistency semantics over the distributed cluster setup to ease
application development.
\item To take advantage of advances in commodity hardware and the
growth of cloud computing
\cite{DBLP:journals/cacm/ArmbrustFGJKKLPRSZ10}, OLTP systems must
target a hardware ecosystem of shared-memory nodes in a
shared-nothing setup, allowing flexible deployments varying in
performance and cost. OLTP systems must also be
designed to utilize the multi-processor, multi-core processor and
multi-tiered cache architecture in individual cluster nodes to
ensure optimal hardware performance \cite{DBLP:conf/edbt/JohnsonPHAF09}.
\end{enumerate}

\subsection{State of the Art}
In recent times there have been numerous implementations of
systems that can be used by HICCUP applications. There has been
an explosion in the implementation of key-value stores, which
provide varying performance, scalability and availability
characteristics. Amazon Dynamo
\cite{DBLP:conf/sosp/DeCandiaHJKLPSVV07}, Google BigTable
\cite{DBLP:journals/tocs/ChangDGHWBCFG08}, Yahoo! PNUTS
\cite{DBLP:journals/pvldb/CooperRSSBJPWY08}, Amazon HBase and
Cassandra \cite{DBLP:journals/sigops/LakshmanM10}, Amazon S3,
Google Cloud Storage, Windows Azure Storage are some examples of
distributed key-value stores. These systems target commodity
hardware in a cluster-based architecture and provide high availability
and scalability at the cost of lower consistency semantics and a low-level
storage-oriented programming model. In order to build HICCUP
applications using these systems, applications must implement multi-key
transactions and manage the low-level key-value based
infrastructure, which is extremely complex and error prone, thus
limiting ease of development
\cite{DBLP:journals/cacm/Terry13}.

In order to fix these problems, transactional capabilities were
introduced in distributed data stores, such as Google Spanner
\cite{DBLP:conf/osdi/CorbettDEFFFGGHHHKKLLMMNQRRSSTWW12} and Megastore
\cite{DBLP:conf/cidr/BakerBCFKLLLLY11}, or exposed using client libraries \cite{warp,
DBLP:conf/osdi/PengD10}. Megastore \cite{DBLP:conf/cidr/BakerBCFKLLLLY11} provides
full ACID semantics but only within partitions of the data, which
limits the ease of application development and flexible
deployments. Warp \cite{warp} and Percolator
\cite{DBLP:conf/osdi/PengD10} provide multi-key
transactions using client libraries but their storage-oriented
programming model and client-based execution make it difficult to
reason about application performance for varying cluster-based
deployments. 

On the other end, classical RDBMS provide a high-level uniform
shared-memory abstraction, which makes application development
extremely easy. However, these systems face performance
scalability issues. In order to meet these challenges, Shore-MT
\cite{DBLP:conf/edbt/JohnsonPHAF09} and DORA
\cite{DBLP:journals/pvldb/PandisJHA10} target shared-memory
architectures for high performance. Shore-MT optimizes a
classical RDBMS engine by reducing or removing contention points for
multi-core and multi-processor hardware. DORA proposes a novel
methodology to partition data amongst the processing threads and
migrates transactions between threads based on the
partitioning. Although both these systems ease application
development, reasoning about program performance remains
extremely hard. These systems are also not suited for cluster-based
deployment unless a shared-memory layer is built underneath them.

H-Store was one of the
first systems to target cluster-based deployments for OLTP
applications \cite{DBLP:journals/pvldb/KallmanKNPRZJMSZHA08}. In order to provide a high-level relational
abstraction, H-Store is reliant on inferring optimal data
partitioning from application logic to minimize the impact of
distributed transactions \cite{DBLP:journals/pvldb/CurinoZJM10}. This makes it extremely difficult for
application developers to reason about program performance over a
variety of applications and increases the performance reliance of
the system on automatic data \cite{DBLP:conf/sigmod/PavloCZ12}
and code partitioning \cite{DBLP:journals/pvldb/CheungAMM12}, or database administrators
(DBAs). This brings database designers and/or DBAs in the critical path
of performance improvements of varying OLTP applications and
ignores application developers as a valuable resource.

Our approach is to provide a high-level, intuitive programming
model over a cluster-based architecture where the uniform
shared-memory space is partitioned by application logic. This
partitioning is done by the application developers, which allows
them to reason about the performance of their programs and
to improve them. By extending transactions to provide this
abstraction with global ACID semantics, our approach maintains
ease of application development.

\section{Approach}
\label{sec:approach}
In this section, we explain our approach in detail with an
example HICCUP application (the TPC-C benchmark) and demonstrate
how it meets the design requirements outlined in Section
\ref{sec:background}.

\subsection{Logical Partitions}
In a cluster-based architecture, the programming model must
provide an abstraction to balance ease of development and
utilization of cluster resources. We want to explore the middle
ground between the two current dominant approaches by allowing
programmers to reason about the performance of their programs in
the distributed setup while maintaining ease of use. We do so by
introducing the notion of logical partitions. A logical partition
forms the unit of logical code isolation. A
programmer architects his application to consist of multiple
application logic units, which communicate with each other, and
establishes control flow dependencies between them. An
application logic unit must be associated with a logical
partition by the programmer. Communication between the
application logic units is expensive, which encourages the
developers to think about the performance of the constructed
program and to make it more efficient. In order to provide ease of
development, ACID semantics are maintained
globally across logical partitions via transactions, which form
the unit of code construction in this model .

In our model, a program (transaction) begins execution on a
logical partition. During its execution the program can only access data
on that logical partition. If it needs to access data on another
logical partition, it must explicitly do so by invoking a program
on another logical partition which is executed as a
subtransaction \cite[p.~253]{tranpro}. When a
transaction or subtransaction is invoked, the logical partition
where it must be executed must be specified as well. Program code
is pre-compiled on each logical partition and executed within the
database process for performance.

In contrast to partitioning data, our model forces the application developer to partition his
application logic. The programmer must construct
his application in terms of distributed logical partitions by
associating application logic to them. Data partitioning is a
consequence of the code partitioning formulated by the
application developer. Existing OLTP systems either provide the
view of a single data partition or fragmented data partitions, but
not partitioning in terms of application logic. This increases
the reliance of these systems on optimal code and data partition
layout, thus making it hard for application developers to reason
about the performance of a variety of applications using these
systems.

In summary, our model allows the application developer to reason
about program performance in terms of distributed application logic units
by exposing coordination costs between logical partitions. The
database designer can focus on providing efficient mapping of
logical partitions to physical partitions (machines) for optimal
cluster resource utilization. Our model allows ease of
construction and deployment of different HICCUP applications
across various cluster configurations.

\minisec{Modeling hard to partition applications.} The model is
ideally suited for applications where partitioning is implicit in
the application logic. Fortunately, this property holds for a
large class of HICCUP applications. Although applications that
are hard to partition, e.g., social networks would not be a
compelling use case for our model, there is no reason for them
to perform worse using our model compared to RDBMS and key-value
stores. On the contrary, the model provides an analytical
playground to application developers to
experiment with different partition layouts. This allows them to
adopt the appropriate model (relational or key-value stores) for
their application depending on the performance needs and
development costs. The goal of the model is not to infer a
partitioning layout for non-partitioned transactions but to
provide a framework to the application developer to specify
transactions with respect to a partitioning layout.\\

\minisec{System Initialization.} The model deterministically
assigns transactions to logical partitions. Since the model
always starts with an initial empty database state, the user
application must provide initialization transactions to create an
initial database state. Since those transactions would be
created using the proposed model, data distribution would always
be deterministic and unambiguous. It is important to mention 
that command logging to provide durability can be interpreted as
an automatic program to generate a given desirable initial database
state. This highlights the focus of the approach to logically
partition code and not data.

\subsection{TPC-C Example}
\begin{figure}[t]
\usebox\bbox
\caption{New\_order transaction}
\label{fig:tpcc:neworder}
\end{figure}

\begin{figure}[t]
\usebox\xbox
\caption{Transaction to read and update stock information}
\label{fig:tpcc:updatestock}
\end{figure}

\begin{figure}[t]
\usebox\ybox
\caption{Mapping transactions to logical partitions}
\vspace{-3ex}
\label{fig:tpcc:mapping}
\end{figure}

The ease of partitioning application logic will determine the
ease of applicability of this approach. HICCUP applications that
have simple application logic are well suited for this approach
as we demonstrate by explaining our approach using the
$new\_order$ transaction class from the TPC-C benchmark as an
example. We will not go into the detailed explanation of the
benchmark which is available in its
specification\footnote{\url{http://www.tpc.org/tpcc/spec/tpcc_current.pdf}}.

We begin by expressing the $new\_order$ application logic without any
notion of partitioning. This is followed by the transformation of the
program into the proposed model by partitioning the $new\_order$
transaction by warehouses. Finally, we show how developers can
reason about the performance of this transaction and rewrite it
in order to make it more efficient.

\subsubsection{New\_Order Application Logic}
The pseudocode for the $new\_order$ transaction is outlined in Figure
\ref{fig:tpcc:neworder}. The pseudocode preserves the data
dependencies in the transaction class, but abstracts away other
details of the implementation including data types. In the
transaction logic, the order is first sanity checked and an order
id is generated. For each item in the order, the amount for the
number of ordered items is computed. The stock of the supplier
warehouse is then updated and an order line is inserted for each item
along with the district information of the stock. Finally, the
total amount to be paid by the customer is computed and returned
by the transaction. It is important to note that the fields
retrieved by the $get\_dist\_info\_stock$ method from the
stock relation, which are used to insert a row in the $order\_line$
relation, are never updated by the $update\_stock$ method. The
price for each item is present in the item relation, which is
never updated and never grows as more warehouses are added.

The TPC-C benchmark was designed to scale in the number of
warehouses, i.e., with additional warehouses, there are more
customers who request greater number of $new\_order$
transactions. Warehouses form the unit of distribution and
scale-up in the benchmark. Many HICCUP applications have similar
properties where a distribution unit leads to growth in both data
and client requests and importantly, application logic is mostly
affine to the distribution unit. \textit{Existing programming models fail
to take advantage of this property}. Our approach, which provides a
high-level, intuitive, partitioned programming model with ACID
guarantees, is ideally suited for this class of applications. 

\subsubsection{Code Partitioning by warehouses} 
\begin{figure}[t]
\usebox\dbox
\caption{New\_order transaction partitioned by warehouses}
\vspace{-3ex}
\label{fig:tpcc:neworderpar}
\end{figure}

In the TPC-C benchmark, a warehouse is the unit of logical
partitioning. This follows intuitively because transaction logic
is warehouse-affine and both data and number of client
transactions increase with additional warehouses. In order to allow orders to be
entirely supplied by a customer warehouse, each warehouse
replicates the item relation. 

Under this logical partitioning model, the new\_order transaction
code is mostly local to the customer warehouse (w\_id). In order
to execute the $get\_dist\_info\_stock$ and $update\_stock$ methods,
the transaction needs to access a remote partition, since these
methods operate on the supplier warehouse. We need to group these methods
under a $new\_order\_update\_stock$ transaction as shown in Figure
\ref{fig:tpcc:updatestock}, so that they can be invoked on the
remote supplier warehouse logical partition.

In the partitioned model, each logical partition is identified by
a unique logical partition identifier. Without loss of
generality, the logical partitions can be identified by unique
natural numbers. When a transaction is invoked, the logical
partition where it would be executed must be specified, which
establishes the relationship between the application logic unit
and a logical partition in the model. This is done by using a
mapping function that operates on the call parameters or any
available context and returns an appropriate partition
identifier. Each transaction class declares the mapping function
it uses and defines it; however, all mapping functions output
partition numbers in the same domain of logical partitions. For
our example, warehouse id can be used as the logical partition
identifier. The mapping function is declared and defined as shown
in Figure \ref{fig:tpcc:mapping}. Now, a $new\_order$ transaction
can be invoked as: \\
\textbf{EXEC $new\_order$ (w\_id, ...) ON PARTITION (w\_id)}

Finally, we re-partition the original $new\_order$ transaction as
shown in Figure \ref{fig:tpcc:neworderpar}, so that it is aware
of logical partitions by invoking $new\_order\_update\_stock$ on remote
supplier warehouse logical partitions. Since transactions
guarantee full ACID semantics even across logical partitions, the
programmers do not have to worry about consistency issues. 

\subsubsection{Further Optimizations}
\begin{figure}
\usebox\cbox
\caption{New\_order transaction with further optimizations}
\label{fig:tpcc:neworderopt}
\end{figure}

A closer look at the partitioned $new\_order$ transaction shown
in Figure \ref{fig:tpcc:neworderpar} exposes several performance
issues owing to control flow dependencies upon remote transaction
results. There is a dependency on the results from the remote
$new\_order\_update\_stock$ transaction in order to insert order
line entries. Since the item relation is already replicated on
each logical partition, the amount can be computed locally. The
real dependency of order line entries is to get the district
information of the stock from the remote supplier warehouse. The
district information fields are never updated so these fields can
be replicated on each logical partition. Then, the
$new\_order\_update\_stock$ transaction can be modified so that
it does not have to return a result and the $new\_order$
transaction can be rewritten as shown in Figure
\ref{fig:tpcc:neworderopt}. In order to invoke subtransactions
in parallel, we provide \textbf{PARALLEL EXEC} construct. In
order to use the construct, the transaction to be executed and
the iterator to be used must be specified. Using this construct
in Figure \ref{fig:tpcc:neworderopt}, ``EXEC
$new\_order\_update\_stock$ (order) ON PARTITION (s\_id)" is
invoked in parallel for all iterator elements (s\_id) in
order.supplier\_w\_id list.

Since $new\_order$ transactions constitute $\sim$43\% of the
workload mix, this optimization might be worthwhile. As the
number of warehouses increase, however, the size of the replicated stock
relation with the district information fields increases as well. For a
large number of warehouses, this optimization option may thus not
be viable. Nevertheless, it is important to note that the model
allowed the application developer to reason about the performance
of the developed programs ($new\_order$ programs) and improve
them using the mentioned optimizations. Consequently, OLTP systems
exposing a logical partitioning model can provide performance
guarantees over varying classes of applications.

\subsubsection{Summary}
In a cluster-based setup, it is important to separate the
concerns of writing efficient programs and ensuring their
efficient execution by mapping to physical machines. This is done
by exposing a programming model based on logical partitioning,
which allows the application developers to reason about program
performance. HICCUP applications, which have an implicit
distribution unit, can be intuitively modeled in the proposed
programming model as shown in the TPC-C example. Strong global
consistency semantics ensure ease of program development.

\section{Challenges and Roadmap}
\label{sec:roadmap}
In this section, we outline the challenges that need to tackled
during the course of the project. We also provide some initial
thoughts on how to address them.
\subsection{Physical Implementation and Evaluation}
\minisec{Logical to Physical Mapping.} In order to deploy a set
of logical partitions over a set of physical partitions (physical
machines with varying hardware characteristics), the logical
partitions need to be mapped to the set of physical
partitions. Each physical partition needs to run a partition
executor that is responsible for handling transactions. This
gives rise to the following sub-challenges, which we plan to
investigate: (1) How to reuse a transaction processing system
which meets performance demands on modern main-memory multi-core
machines and extend it with the logical programming abstraction?
(2) How to map a set of logical partitions to a set of physical
partitions to ensure optimal resource usage of the cluster?

In order to meet the first challenge, we plan to use a
state-of-the-art transaction processing system designed for
main-memory multi-core machines
\cite{DBLP:conf/sosp/TuZKLM13}, and extend it to operate in a
distributed setup to provide the logical partitioning
abstraction. In order to meet the second challenge, we plan to
investigate cost models that attempt to approximate transaction
execution costs in a physical cluster setup. This involves
identifying and characterizing metrics in the cost model. The
cost model can then be used to produce an initial mapping, which
can be changed if workload characteristics change. We also plan
to investigate transaction scheduling strategies without
violating performance constraints, which would help in better
resource utilization.

\minisec{Local Concurrency Control and Global Commit.} One of the
simplest models to run multiple transactions is to run them
serially, which is what H-Store
advocates \cite{DBLP:journals/pvldb/KallmanKNPRZJMSZHA08}. This simple
model will not work for distributed transactions unless a global
schedule is pre-decided
\cite{DBLP:conf/sigmod/ThomsonDWRSA12}. Running transactions
serially would be a waste of compute power in multi-core
architectures of today. There is also a need to hide memory
latencies, network latencies and commit latencies. However,
multi-threading with pessimistic concurrency control leads to
contention bottlenecks in the lock manager
\cite{DBLP:conf/edbt/JohnsonPHAF09} and multi-phase commits, which hurt
performance.

In a partitioned model, distributed optimistic concurrency
control holds promise owing
to its parallel nature, lack of long critical sections and
multi-phase commits \cite{DBLP:journals/tods/Herlihy90}. It provides a global commit order in the
distributed setup using local commit ordering decisions without
necessitating a multi-phase protocol and hindering local
transactions. We want to investigate the performance of a
distributed optimistic concurrency control mechanism in order to
support the transactional semantics of our approach. We also want
to investigate how transaction aborts hurt performance
under contention, and whether hybrid concurrency control and/or
program repartitioning would improve system performance.

\minisec{Evaluation.} In order to evaluate the system
implementation, we plan to measure the scalability in transaction
throughput, latency, abort rates and cost per transaction with
growing data and number of clients. We are particularly
interested in evaluating these metrics for various physical
machine configurations and logical-to-physical partition
mappings. We plan to evaluate the system using the TPC-C
benchmark. We are also interested in looking at other
non-standard application benchmarks, which could provide clues on
usability of the proposed model. We plan to compare the results
with that of H-Store, a commercial RDBMS and Silo to characterize
the system performance and understand the architecture design
trade-offs.

\subsection{Cloud Integration}
The cloud computing infrastructure provides a flexible ecosystem
for deployment of the system. In order to target the cloud
computing infrastructure and satisfy varying HICCUP application
requirements, we plan to investigate the necessary set of
self-tuning tools to make the system cloud-ready. Deployment
advisors that self-manage diverse, changing resources
\cite{DBLP:conf/middleware/YinSCBR09}, adapt to the cloud
infrastructure \cite{DBLP:journals/pvldb/ZouBSDG12}, have
received a lot of attention lately. We plan to develop a
deployment advisor that generates an optimal cluster
configuration in the cloud from an application's requirements
(e.g., throughput, latency, cost constraints and goals) and
programs by constructing and solving an optimization problem.
We plan to investigate techniques to allow the deployment advisor
to adapt to the variance and workload mix in the cloud infrastructure.

\section{Conclusion}
Designing main-memory OLTP systems that provide high performance,
scalability and ease of development over a variety of HICCUP
applications is an open challenge. In this paper, we identify the
need to allow application developers to reason about the
performance of their programs while maintaining ease of use. We
propose an extension to the transactional abstraction to provide
a high-level, intuitive, distributed logical programming model
with strong global consistency semantics to the application
developer. We have also shown how the model can be intuitively
used by the application developer by using TPC-C as an
example. We plan to evaluate the potential of this approach to
meet the requirements of a variety of HICCUP applications.
\bibliographystyle{vldb}
\bibliography{paper}
\balance
\end{document}